\documentclass[twocolumn]{article}
\usepackage{psfig}
\begin{document}

%\title{\hfill \huge \bf letters to Nature}
%\maketitle

\noindent{\Large \bf 
Measurement of stellar age from uranium decay}

\bigskip

\noindent { 
R. Cayrel$^1$, V. Hill$^2$, T. C. Beers$^3$, B. Barbuy$^4$, 
M. Spite$^5$, F. Spite$^5$, B. Plez$^6$, J. Andersen$^7$, 
P. Bonifacio$^8$, P. Fran\c{c}ois$^9$,
P. Molaro$^8$, B. Nordstr\"{o}m$^{7,10}$ \& F. Primas$^2$}

\bigskip

\noindent{\it \small
$^1$Observatoire de Paris-Meudon, DASGAL, F 75014 Paris, France\\
$^2$European Southern Observatory, D 85748 Garching b. M\"unchen, Germany\\
$^3$Michigan State University, East Lansing, Michigan 48824, USA\\
$^4$Universidade de S\~ao Paulo, S\~ao Paulo, BR 01060-970, Brazil\\
$^5$Observatoire de Paris-Meudon, DASGAL, F 92195 Meudon Cedex,
France\\
$^6$GRAAL, Universit\'e Montpellier-2, F 34095 Montpellier Cedex, France\\
$^7$University of Copenhagen, Astronomical Observatory, DK 2100, Copenhagen,
Denmark\\
$^8$Osservatorio Astronomico di Trieste, I-34131 Trieste, Italy\\
$^9$European Southern Observatory, Casilla 19001, Santiago 19, Chile\\
$^{10}$Lund Observatory, Lund University, S 22100 Lund, Sweden\\
}

\bigskip

{\bf The ages of the oldest stars in the Galaxy indicate when star formation
began, and provide a minimum age for the Universe. Radioactive dating of
meteoritic material$^1$ and stars$^2$ relies on comparing the present abundance
ratios of radioactive and stable nuclear species to the theoretically
predicted ratios of their production. The radioisotope $^{232}$Th (half-life
14 Gyr) has been used to date Galactic stars$^{2-4}$, but it decays by only a
factor of two over the lifetime of the Universe. $^{238}$U (half-life
4.5 Gyr) is 
in principle a more precise age indicator, but even its strongest spectral
line, from singly ionized uranium at a wavelength of 385.957 nm, has
previously not been detected in stars$^{4-7}$. Here we report a measurement of
this line in the very metal-poor star CS31082-0018, a star which is strongly
overabundant in its heavy elements. The derived uranium abundance, 
$\log \rm (U/H)=-13.7\pm 0.14\pm 0.12$ yields an age of $12.5\pm 3$ Gyr,
though this is still model dependent. The observation of this
cosmochronometer gives the most direct age determination of the Galaxy.
Also, with improved theoretical and laboratory data, it will provide a
highly precise lower limit to the age of the Universe.}

We are currently undertaking a large programme of high-resolution
spectroscopy -at the ESO Very Large Telescope and UVES$^{9}$ spectrograph- of
extremely metal-poor stars selected primarily from the Ca II HK survey of
ref.8. The star CS31082-001 (V-band magnitude V = 11.7) has been identified
as very metal-poor (T.C.B. et al., manuscript in preparation). One of us
(the observer, V.H.) has noted that this star is similar to another well
studied very metal-poor star, CS22892-052$^{6,10,11}$, in that it exhibits a
large enhancement (relative to iron) of elements formed by the
neutron-capture r-process. High-resolution spectroscopic observations of
CS22892-052 had allowed a precise measurement of $^{232}$Th from the 401.9nm
line, but based on the abundance of this slowly decaying species, the age of
this star remained fairly uncertain: $15.6\pm 4.6$ Gyr (ref.4).

Moreover, the newly discovered star CS31082-001 exhibits considerably less
contamination of the atomic line spectrum by molecular bands of CH and CN
than does CS22892-052. The Th II line at 401.9 nm is also unusually strong
and clear of contaminants; in fact, a total of 14 Th II lines are detected,
11 of which were selected for measurement. Only the 401.9nm line has
previously been accurately measured in a stellar spectrum, and 10 additional
lines in CS31082-001 appear to be first detections. More importantly, the
strongest U II line is clearly detected (Fig. 1). We note that no lines are
seen at this wavelength in the stars HD1154444 or HD122563, which have
atmospheric parameters and iron abundance similar to those of CS31082-001,
removing any doubt that the line we see is indeed due to U II. To illustrate
the quality of our data, a one-hour integration at a resolving power of R =
70,000 yields typical signal-to-noise (S/N) ratios of 150 at 385 nm, and
250-300 at 650 nm. The region of the U II line is currently covered by eight
such individual spectra. Other lines of U II are below 1 m\AA~ in equivalent
width in our spectra, and blended with stronger lines.

\begin{figure}[h]
\psfig{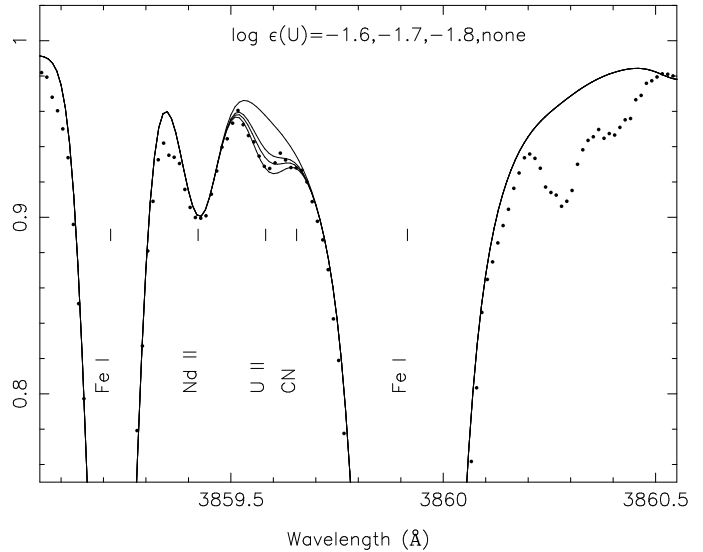}
\caption{The spectrum of CS31082-001 around the U II line at
385.959 nm. The synthetic spectra (solid lines) were computed with the
stellar atmospheric parameters given in the text, and for the three
abundances indicated, adopting an oscillator strength $f=0.053$ for
the line (ref.13). The observed spectrum (data points) was obtained
in four hours for a total S/N ration of 300.}
\end{figure}

Using available visual and infrared photometry and the detailed
spectroscopic constraints from our own observations, the best model
atmosphere for CS31082-001 has the parameters effective temperature T$_{eff}$ =
4,825 K, gravity [g(cm s$^{-2}$)] $\log g = 1.5\pm 0.2$, and microturbulence
$\xi_{micro} = 1.8\pm 0.2 \rm km s^{-1}$. Thermal non-equilibrium
effects should 
be unimportant for Th II and U II, as these elements are virtually fully
ionized and the observed transitions come from the ground level or from a
very low excitation level. We then derive the following abundances (on the
scale where $\log \epsilon$(H) = 12): 
$\log \epsilon \rm (Fe) = 4.60\pm 0.06$,
$\log \epsilon \rm (Os) = 0.49\pm 0.10$;
$\log \epsilon \rm (Ir) = 0.40\pm 0.12$, 
$\log \epsilon \rm (Th) = -0.96\pm 0.03$, and 
$\log \epsilon \rm (U) = -1.70\pm 0.10$. 
The quoted errors are the standard deviation on the mean
abundance from at least three lines when available, otherwise they are
estimates from the S/N ratio of the spectrum. They do not include systematic
errors on oscillator strengths, which are extremely difficult to assess. For
U we recomputed the partition functions of U I, U II and U III from the
tables of energy levels in ref.12, and adopted the oscillator strength from
the laser-induced fluorescence measurements of ref.13.

The iron abundance of CS31082-001 is about 1/800 of that of the Sun, while
the heaviest detected stable elements, Os and Ir, are about 1/9 as abundant
as in the Sun. CS31082-001 is exceedingly rich in spectral lines from most
or all of the heavy elements, and a detailed discussion of the individual
elements and the process(es) by which they were formed will be reported
elsewhere (R.C. et al., manuscript in preparation).

All the heavy elements in a star as metal-poor as CS31082-001 were likely to
have formed during a very short interval early in the history of the Galaxy
itself$^{14,15}$. The abundance ratio, $r$, between a radioactive and a stable
comparison species observed $\Delta t$ Gyr after production is then 
$\rho =\rho_0\bigl[\exp(-ln2\times\Delta t/t_{1/2})\bigr]$, 
where $\rho_0$ is the initial
production ratio and $t_{1/2}$ is the half-life of the radioactive isotope.

In order to determine an absolute radioactive decay age, the initial
production ratio is needed. Such ratios are predicted by theoretical
nucleosynthesis models, validated by requiring that the predicted abundance
ratios for all stable r-process elements in a neighbouring range of mass
numbers should be identical to those observed in the star and in the Solar
System$^{4,16}$. The accuracy of a predicted ratio depends both on the adopted
nuclear physics model and on the degree to which different production sites
yield similar results. The latter has generally been found to be the case
for previously studied r-process enhanced stars, but further work is needed
to verify it in detail for CS31082-001.

Both theoretical and observational uncertainties should be minimal for
elements as close as possible to each other in atomic number. Thus, Os and
Ir should be the best stable reference elements in CS31082-001 for the
radioactive species $^{232}$Th and $^{238}$U (any $^{235}$U has decayed to
insignificance long ago). From the half-lives of $^{232}$Th and
$^{238}$U (14.05 and 4.468 Gyr, respectively), $\Delta t$ as a
function of the logarithmic decay of Th and U is:
\begin{equation}
\Delta t = 46.7 \bigl[\log ({\rm Th}/r)_0 - \log ({\rm Th}/r)_{obs}\bigr]
\end{equation}
\begin{equation}
\Delta t = 14.8 \bigl[\log ({\rm U}/r)_0 - \log ({\rm U}/r)_{obs}\bigr]
\end{equation}
\begin{equation}
\Delta t = 21.8 \bigl[\log ({\rm U}/Th)_0 - \log ({\rm U}/Th)_{obs}\bigr]
\end{equation}
where $\Delta t$ is expressed in Gyr, $r$ is a stable third-peak r-process
element (here Os or Ir), and the terms such as (Th/$r$)$_0$ are the initial
production ratios for each pair of species.

Equations (1) and (2) highlight the importance of adding U to the battery of
Galactic chronometers: an error of 0.1 dex in a Th abundance propagates as a
time equal to the age of the Solar System, neglecting other sources of
error. But equation (3) shows that U and Th can be used in concert, with
little loss in formal precision but with important overall advantages: the
initial production ratio (U/Th)0 is in principle much less affected$^{17}$ by
theoretical uncertainties than any of the individual (U/$r$)$_0$ because of the
proximity of the two elements in their mass numbers. We note that the
observational accuracy of (U/Th) is better than for U or Th alone, because
errors coming from the choice of model atmosphere parameters largely cancel
out.

The observed value of log(U/Th) in CS31082-001 is $-0.74\pm 0.15$. Here
our error estimate includes 0.1 dex reflecting astrophysical observational
errors, and 0.12 dex for the uncertainty on the oscillator strength of the
U II 385.96-nm line. We only have to introduce an estimate of log(U/Th)$_0$ in
equation (3) to derive the age of CS31082-001. This is done in Table 1,
where we have given the reference used for the value of the production
ratio. We have also explored the use of the ratio of U to stable elements.
As explained above, it is safest to choose a reference element as heavy as
possible (that is, close in mass number to U and Th). Table 1 gives the
corresponding results using production ratios for Os and Ir from ref.4. Any
age between 11.1 and 13.9 Gyr is compatible with the various determinations
associated with their error bars. We consider the median value 12.5 Gyr as
our best present estimate for the age of CS31082-001, with a conservative
standard error of 3 Gyr. When increased by 0.1-0.3 Gyr (refs 14, 15), these
values give the age of the Galaxy, which is in turn a lower limit to the age
of the Universe.

\begin{table}\caption{Ages derived for CS31082-001 as a function of
production ratios}
\begin{tabular}{l@{ }c@{ }c@{ }c@{ }c}
\hline
\footnotesize Element pair& 
\footnotesize $\log$(prod. ratio)&
\footnotesize Ref &
\footnotesize $\log$(obs. ratio)&
\footnotesize Derived age (Gyr)\\
\hline
U/Th & -0.255 & 4& -0.74$\pm 0.15$ & 10.6$\pm 3.3$\\
U/Th & -0.10  &17& -0.74$\pm 0.15$ & 14.0$\pm 3.3$\\ 
U/Os & -1.27  & 4& -2.19$\pm 0.18$ & 13.6$\pm 2.7$\\ 
U/Ir & -1.30  & 4& -2.10$\pm 0.17$ & 11.8$\pm 2.5$\\ 
\hline
\end{tabular}
\end{table}

The accuracy of this uranium dating technique is at present limited by
incomplete knowledge of a few critical physical data, in particular
oscillator strengths and production ratios of the elements produced by the
r-process. Further laboratory and theoretical work should enable progress on
both these issues, as should the detection of additional stars with enhanced
abundances of r-process elements, similar to CS31082-001. Such work, already
in progress, should allow the full potential of the uranium chronometer to
be realized.

\medskip

{\footnotesize \it Received 9 November 2000;accepted 29 December 2000}

\medskip

{\bf References} \\ \small
  1. Tilton, G. R. in Meteorites and the Early Solar System (eds Kerridge,
    J. F. \& Matthews, M. S.) 249-258 (Univ. Arizona Press, Tucson, 1988).\\
  2.Butcher, H. R. Thorium in G-dwarfs as a chronometer for the Galaxy.
    Nature 328, 127-131 (1987).\\
  3. Fran\c{c}ois, P., Spite, M. \& Spite, F. On the galactic age problem:
    Determination of the [Th/Eu] ratio in halo stars. Astron. Astrophys.
    274, 821-824 (1993).\\
  4. Cowan, J. J. et al. r-process abundances and chronometers in metal-poor
    stars. Astrophys. J. 521, 194-205 (1999).\\
  5. Westin, J., Sneden, C., Gustafsson, B. \& Cowan, J. J. The r-process
    enriched low-metallicity giant HD 115444. Astrophys. J. 530, 783-799
    (2000).\\
  6. Sneden, C. et al. Evidence of multiple r-process sites in the early
    Galaxy: New observations of CS 22892-052. Astrophys. J. 533, L139-L142
    (2000). \\
  7. Sneden, C. \& Cowan, J. J. The age of the Galaxy from thorium
    cosmochronometry. Rev. Mex. Astron. Astrophys. (in the press); also
    preprint astro-ph/0008185 at http://xxx.lanl.gov (2000).\\
  8. Beers, T. C., Preston, G. W. \& Shectman, S. A. A search for stars of
    very low metal abundance. II. Astron. J. 103, 1987-2034
    (1992).\\
  9. D'Odorico, S. et al. The performance of UVES, the echelle spectrograph
    for the ESO VLT, and highlights of the first observations of stars and
    quasars. Proc. SPIE 4005, 121-130 (2000).\\
 10. McWilliam, A., Preston, G. W., Sneden, C. \& Searle, L. Spectroscopic
    analysis of 33 of the most metal-poor stars. II. Astron. J. 109,
    2757-2799 (1995).\\
 11. Sneden, C. et al. The ultra-metal-poor, neutron-capture-rich giant star
    CS 22892-052. Astrophys. J. 467, 819-840 (1996).\\
 12. Blaise, J. \& Wyart, J.-F. Tables Internationales de Constantes
    S\'electionn\'ees Vol. 20, Energy Levels and Atomic Spectra of the
    Actinides (Univ. Pierre et Marie Curie, Paris, 1992).\\
 13. Chen, H.-L. \& Borzileri, C. Laser induced fluorescence studies of U II
    produced by photoionization of uranium. J. Chem. Phys. 74, 6063-6069
    (1981).\\
 14. Argast, D., Samland, M., Gerhard, O. E. \& Thielemann, F.-K. Metal-poor
    halo stars as tracers of ISM mixing processes during halo formation.
    Astron. Astrophys. 356, 873-887 (2000).\\
 15. Chiappini, C., Matteucci, F., Beers, T. C. \& Nomoto, K. The earliest
    phases of Galaxy formation. Astrophys. J. 515, 226-238 (2000).\\
 16. Pfeiffer, B., Kratz, K. -L. \& Thielemann, F.-K. Analysis of the
    solar-system r-process abundance pattern with the new ETFSI-Q mass
    formula. Z. Phys. A 357, 235-238 (1997).\\
 17. Goriely, S. \& Clerbaux, B. Uncertainties in the Th cosmochronometry.
    Astron. Astrophys. 346, 798-804 (1999).\\
 18. Fouqu\'e, P. et al. An absolute calibration of DENIS (Deep Near Infrared
    Southern Sky Survey). Astron. Astrophys. Suppl. 141, 313-317 (2000).\\
\normalsize

\medskip

{\bf Acknowledgements.}
 We thank the ESO staff for assistance during the
observations, J.-F. Wyart and L. Tchang-Brillet for helping us to obtain the
best current atomic data for Th and U, and G. Simon for providing the
unpublished infrared colours for CS 31082-001 from the DENIS$^{18}$ survey.
Partial financial support for this work was obtained from the US NSF (to
T.C.B.), and from the Carlsberg Foundation and Julie Damms Studiefond (to
J.A. and B.N.).

\medskip

{\it Correspondence should be addressed to R.C. (e-mail:
Roger.Cayrel@obspm.fr).}

\end{document}